\documentclass[aps,preprint,nofootinbib]{revtex4}%
\usepackage{amsfonts}
\usepackage{amsmath}
\usepackage{amssymb}
\usepackage{graphicx}%
\providecommand{\U}[1]{\protect\rule{.1in}{.1in}}

\begin{document}
\title{Translational invariance of the Einstein-Cartan action in any dimension}
\author{N. Kiriushcheva}
\affiliation{Department of Applied Mathematics, University of Western Ontario, London, Canada}
\email{nkiriush@uwo.ca}
\author{S.V. Kuzmin}
\affiliation{Faculty of Arts and Social Science, Huron University College and Department of
Applied Mathematics, University of Western Ontario, London, Canada}
\email{skuzmin@uwo.ca}
\date{\today}

\begin{abstract}
We demonstrate that from the first order formulation of the Einstein-Cartan
action it is possible to derive the basic differential identity that leads to
translational invariance of the action in the tangent space. The
transformations of fields is written explicitly for both the first and second
order formulations and the group properties of transformations are studied.
This, combined with the preliminary results from the Hamiltonian formulation
(arXiv:0907.1553 [gr-qc]), allows us to conclude that without any
modification, the Einstein-Cartan action in any dimension higher than two
possesses not only rotational invariance but also a form of
\textit{translational invariance in the tangent space}. We argue that
\textit{not }only a complete Hamiltonian analysis can unambiguously give an
answer to the question of what a gauge symmetry is, but also the pure
Lagrangian methods\ allow us to find the same gauge symmetry from the
\textit{basic} differential identities.

\end{abstract}
\maketitle

\section{Introduction}

From the Hamiltonian analysis of the first order form of the Einstein-Cartan
(EC) action without coupling to matter and using the N-beins and connections
as independent variables, it is clear that the only \textit{gauge} invariance
that follows from the first class constraints, in addition to the rotational
(\textquotedblleft Lorentz\textquotedblright) invariance in the tangent space,
is some form of translational invariance also in the tangent space (for some
preliminary results using the Hamiltonian analysis see \cite{Report}). In all
dimensions higher than two this invariance must possess a parameter with an
internal index that corresponds to either the translational part of the
Poincar\'{e} symmetry (as in the three dimensional case \cite{3D}) or a
Poincar\'{e} symmetry with a \textquotedblleft more general group structure
that the original Poincar\'{e} group\textquotedblright\ \cite{Hehl}. In a flat
spacetime limit this \textquotedblleft more general group
structure\textquotedblright\ is equivalent to the Poincar\'{e} group,
providing a natural generalization of Poincar\'{e} symmetry to curved
spacetime \cite{Trautman}. In supergravity a translational invariance in the
tangent space was used for the first time by Breitenlohner
\cite{Breitenlohner} to avoid \textquotedblleft on-shell\textquotedblright%
\ closure of the commutator algebra. The paper which is most closely related
to our analysis (to both Hamiltonian \cite{Report, 3D} and Lagrangian
analysis) is one by Teitelboim \cite{Teitelboim}, where translational and
rotational invariances in the tangent space were discussed and the structure
of the algebra of constraints for both gravity and supergravity, as well as
the resultant transformations of fields, were predicted using very general
observations. The explicit form of either the constraints or the structure
functions were not given in \cite{Teitelboim}, but our Hamiltonian
analysis\ of the EC action leads exactly to such a structure.

However a different opinion commonly appears in the literature about the
ordinary EC action (without considering any generalization and
supersymmetrizations). For example, in \cite{Blag} it is stated that the EC
action \textquotedblleft is invariant under Lorentz rotations and
diffeomorphisms, but \textit{not under translations [italic of
M.B.]\textquotedblright,\ }and that only in three dimensions is the EC action
a gauge theory with both rotational and translational invariances. According
to \cite{Banados}, \textquotedblleft... a closer analysis shows that only
Lorentz rotations are symmetries of the action: the translational part of the
Poincar\'{e} group does not leave the action invariant\textquotedblright\ and
\textquotedblleft...gravity is not a gauge theory...\textquotedblright\ 

The statements that there is no translational invariance in dimensions higher
than three have appeared in the literature without proof, but quite often the
reference on Witten's work \cite{Witten} is used. However, Witten did not draw
this conclusion, only stating: \textquotedblleft we cannot hope that
four-dimensional gravity would be a gauge theory in this
sense\textquotedblright\ (in the sense that it would be similar to the
Chern-Simons action). We do not see how this statement was generalized to
saying that EC gravity is not a gauge theory in dimensions higher than three
\cite{Banados} or that it is a theory with diffeomorphism as a \textit{gauge}
symmetry \cite{Gren, Carlip, Banados, Ali} (with transformations in the
external space), in place of translational invariance in the internal space,
which is a part of Poincar\'{e} or generalized Poincar\'{e}
symmetry.\footnote{In some papers Lorentz symmetry plus diffeomorphism are
erroneously called the Poincar\'{e} gauge symmetry (see, for example,
\cite{Gren, Ali}).} In many papers the authors switch without justification
from the translational invariance in the tangent space (which naturally arises
from the constraint structure in the first steps of the Hamiltonian
formulation \cite{3D, Report}, as well as from the basic differential
identities of the Lagrangian formulation (see next Section)) to diffeomorphism
invariance (translation in the external space) in an attempt to try to obtain
the \textquotedblleft expected\textquotedblright\ result, for further
\textquotedblleft convenience\textquotedblright\ (see, e.g., \cite{Kibble,
Gren, Nicolic, Banados, Ali}; this is even done for the EC action in three
dimensions \cite{Witten, Carlipbook}). The EC action is manifestly
diffeomorphism invariant but this is not necessary a gauge invariance if by
\textquotedblleft gauge invariance\textquotedblright\ we mean a transformation
generated by the first class constraints of the theory. It has been stated
that there exist various ways to define the constraints of tetrad gravity
which lead to different gauge transformations \cite{Nelson}. However, at the
Hamiltonian level, such changes are possible only if one makes non-canonical
transformations of variables. This erroneously assumes that a unique
characteristic of a theory, namely the gauge invariance which follows from the
first class constraints of the theory, can be chosen according to one's
preference.\textit{ }

All symmetries of the EC action are encoded in it (as in any field theory) and
should be found without any \textit{a priori }knowledge of what symmetry we
expect. One of many symmetries of EC action is translational invariance
(contrary to what is stated in some references) and we will show that this
symmetry follows from the Lagrangian in the most natural way, just as the
rotational symmetry can be derived from the Lagrangian, and there is no need
to modify the action to have such an invariance. The question whether
translation is a gauge symmetry which follows from the first class constraints
of the EC action or it \textquotedblleft is not a gauge
theory\textquotedblright\ \cite{Banados} can be answered by using Hamiltonian methods.

In the Hamiltonian approach to constrained (singular) systems \cite{Diracbook}
there is the well-defined criterion for a theory being a gauge theory: if
first class constraints are present then the theory possesses a gauge symmetry
that can be found uniquely from the Poisson brackets (PBs) algebra of these
first class constraints. To find this gauge symmetry and the associated
transformations of the fields, second class constraints have to be eliminated
(the Hamiltonian reduction), and by applying some standard methods (e.g. the
oldest one due to Castellani \cite{Castellani}), gauge transformations of the
fields can be obtained. Of course, the Hamiltonian procedure destroys manifest
covariance, especially after elimination of the second class constraints
(which are always present in first order formulations), and if the gauge
transformations are covariant then restoring their covariant form requires
labourious calculations using solutions of second class constraints. (For some
examples see \cite{AOP}, \cite{3D}, and \cite{Report}.)

In contrast, if one is not interested in obtaining the reduced Hamiltonian by
elimination of non-physical degrees of freedom, the explicit form of the
constraints, the algebra of PBs, etc., and is only interested in finding all
symmetries of an action without determining which of the symmetries is a gauge
symmetry generated by the first class constraints and which is not, then
Lagrangian methods are preferable as they retain manifest covariance making
calculations much simpler and more transparent. In some works (e.g. see a
recent article of Samanta \cite{Samanta} and references therein), there are
examples of such \textquotedblleft Lagrangian\textquotedblright\ approaches.
However, in \cite{Samanta}, covariance is needlessly broken and the approach
presented is just a modification of the Hamiltonian method of Castellani
\cite{Castellani}. Such a modification is unnecessary, especially because in
applications of this \textquotedblleft Lagrangian\textquotedblright\ method
differential identities (DIs) are taken for granted \cite{Samanta} and they
are always in covariant form. If DIs are known, the corresponding
transformations of the fields are simply obtained. A method of finding a
general DI does not exist in these approaches; artificially destroying
covariance in such \textquotedblleft Lagrangian\textquotedblright\ methods
makes this task impossible without \textit{a priori} knowledge of the
symmetries of the theory. Indeed, if one uses known symmetries of the theory
in conjunction with the corresponding DIs, one can obtain the associated
Noether current in the theory \cite{Noether}.

In model building many approaches are based on postulating particular
symmetries and constructing actions that support them. Our goal is different:
using the EC action as an example, we demonstrate that there is an algorithm
for building differential identities that allows us to find all symmetries of
an action without \textit{a priori} knowledge of what symmetries it has (not
to build an action that should possess some symmetries). We explicitly
demonstrate that statements that translational invariance in dimensions higher
than three is absent are groundless by deriving such transformations for the
EC action. The translational invariance that we find has been observed in
\cite{Hehl, Trautman, Leclerc}.

In this paper, using the purely Lagrangian approach (Section II) we find the
basic DIs which directly lead to both rotational and translational invariances
in the tangent space for the EC action in all dimensions higher than three. In
Section III the group structure of some invariances and their different
combinations are discussed. Using the equivalence of the Lagrangian and
Hamiltonian methods, in our conclusion we address the question
\textquotedblleft what is a gauge symmetry and what is not?\textquotedblright%
\ \cite{Matschull}. We argue that pure Lagrangian methods also can be used to
unambiguously single out the same symmetry that follows from Hamiltonian
methods (i.e. originates with the first class constraints). This is what we
refer to as a gauge symmetry.

\section{Translational invariance of the Einstein-Cartan action}

The EC action in its first order form where N-bein fields $e_{\mu\left(
\alpha\right)  }$ and connections $\omega_{\mu\left(  \alpha\beta\right)
}=\omega_{\mu\left(  \beta\alpha\right)  }$ are independent
variables\footnote{Usually variables $e_{\gamma\left(  \rho\right)  }$ and
$\omega_{\nu\left(  \alpha\beta\right)  }$ are called tetrads and spin
connection, but such names are specialized for $D=4$. As we consider the
Lagrangian formulation in any dimension ($D>2$), we will call $e_{\gamma
\left(  \rho\right)  }$ and $\omega_{\nu\left(  \alpha\beta\right)  }$ N-beins
and connections, respectively.} is (found, e.g., in \cite{Schwinger, CNP})%

\begin{equation}
S\left(  e,\omega\right)  =%
{\displaystyle\int}
d^{D}xL\left(  e,\omega\right)  \label{eqn1}%
\end{equation}
with the Lagrange density%

\begin{equation}
L\left(  e,\omega\right)  =-eA^{\mu\left(  \alpha\right)  \nu\left(
\beta\right)  }\left(  \omega_{\nu\left(  \alpha\beta\right)  ,\mu}%
+\omega_{\mu\left(  \alpha\gamma\right)  }\omega_{\nu~~\beta)}^{~(\gamma
}\right)  =-ee^{\mu\left(  \alpha\right)  }e^{\nu\left(  \beta\right)  }%
R_{\mu\nu\left(  \alpha\beta\right)  }=-\frac{1}{2}eA^{\mu\left(
\alpha\right)  \nu\left(  \beta\right)  }R_{\mu\nu\left(  \alpha\beta\right)
} \label{eqn2}%
\end{equation}
where $e=\det\left(  {e_{\gamma\left(  \rho\right)  }}\right)  $, $D$ is the
dimension of spacetime and%

\[
A^{\mu\left(  \alpha\right)  \nu\left(  \beta\right)  }=e^{\mu\left(
\alpha\right)  }e^{\nu\left(  \beta\right)  }-e^{\mu\left(  \beta\right)
}e^{\nu\left(  \alpha\right)  }=-A^{\mu\left(  \beta\right)  \nu\left(
\alpha\right)  }=-A^{\nu\left(  \alpha\right)  \mu\left(  \beta\right)  }.
\]
The properties of $A^{\mu\left(  \alpha\right)  \nu\left(  \beta\right)  }$
and other similar functions which considerably simplify calculations are
presented in Appendix A.

We also use the Ricci tensor, antisymmetric in $\alpha\beta$ and $\mu\nu$,%

\begin{equation}
R_{\mu\nu\left(  \alpha\beta\right)  }=\omega_{\nu\left(  \alpha\beta\right)
,\mu}-\omega_{\mu\left(  \alpha\beta\right)  ,\nu}+\omega_{\mu\left(
\alpha\gamma\right)  }\omega_{\nu~~\beta)}^{~(\gamma}-\omega_{\nu\left(
\alpha\gamma\right)  }\omega_{\mu~~\beta)}^{~(\gamma}=-R_{\nu\mu\left(
\alpha\beta\right)  }=-R_{\mu\nu\left(  \beta\alpha\right)  }. \label{eqn3}%
\end{equation}
Indices in brackets (..) denote the internal (\textquotedblleft
Lorentz\textquotedblright) indices, and indices without brackets are external
or \textquotedblleft world\textquotedblright\ indices. Internal and external
indices are raised and lowered by the Minkowski tensor $\eta_{(\alpha)(\beta
)}=\left(  -,+,+,...\right)  $ and the metric tensor $g_{\mu\nu}=e_{\mu\left(
\alpha\right)  }e_{\nu}^{\left(  \alpha\right)  }$, respectively. We assume
that the inverse $e^{\mu\left(  \alpha\right)  }$ exists and $e^{\mu\left(
\alpha\right)  }e_{\gamma\left(  \alpha\right)  }=\delta_{\gamma}^{\mu}$,
$e^{\mu\left(  \alpha\right)  }e_{\mu\left(  \beta\right)  }=\delta_{\left(
\beta\right)  }^{\left(  \alpha\right)  }.$ Note that neither the
contravariant N-bein $e^{\mu\left(  \alpha\right)  }$ nor the metric tensor
$g_{\mu\nu}$ are independent variables.

Variation of the action (\ref{eqn1}) with respect to the independent fields is
given by%

\begin{equation}
\delta S=%
{\displaystyle\int}
d^{D}x\left(  E^{\mu\left(  \alpha\right)  }\delta e_{\mu\left(
\alpha\right)  }+E^{\mu\left(  \alpha\beta\right)  }\delta\omega_{\mu\left(
\alpha\beta\right)  }\right)  =0 \label{eqn4}%
\end{equation}
where the Euler derivatives (EDs) are%

\begin{equation}
E^{\tau\left(  \rho\right)  }=\frac{\delta L}{\delta e_{\tau\left(
\rho\right)  }}=-eB^{\tau\left(  \rho\right)  \mu\left(  \alpha\right)
\nu\left(  \beta\right)  }\left(  \omega_{\nu\left(  \alpha\beta\right)  ,\mu
}+\omega_{\mu\left(  \alpha\gamma\right)  }\omega_{\nu~~\beta)}^{~(\gamma
}\right)  , \label{eqn5}%
\end{equation}

\begin{equation}
E^{\tau\left(  \rho\sigma\right)  }=\frac{\delta L}{\delta\omega_{\tau\left(
\rho\sigma\right)  }}=eB^{\nu\left(  \alpha\right)  \mu\left(  \rho\right)
\tau\left(  \sigma\right)  }e_{\nu\left(  \alpha\right)  ,\mu}-eA^{\tau\left(
\rho\right)  \nu\left(  \beta\right)  }\omega_{\nu~~\beta)}^{~(\sigma
}+eA^{\tau\left(  \sigma\right)  \nu\left(  \beta\right)  }\omega_{\nu
~~\beta)}^{~(\rho}. \label{eqn6}%
\end{equation}

Note that the EDs are the result of variation of an action with respect to the
independent fields. In this $e=\det\left(  {e_{\gamma\left(  \rho\right)  }%
}\right)  $ is also the part of the variation, and thus the EDs are different
from equations of motion (see, e.g. \cite{Samanta}) where $e$ can be omitted,
and other rearrangements can be performed such as contraction with fields,
taking different combinations of equations of motion, etc.

Systematic studies of symmetries of any action can be performed by
constructing differential identities which are linear combinations of EDs. DIs
identically equal zero without using solutions of the corresponding equations
of motion (they are valid \textquotedblleft off-shell\textquotedblright). In
general, coefficients of the EDs in such linear combinations are functions of
the fields and/or their derivatives. Such differential identities (if found)
immediately lead to an invariance of the action under the corresponding
transformations of fields. For example, let us say, we construct an identity
which has one external index $I^{\mu}=0$. Multiplying it by a parameter
$\xi_{\mu}$ to form a scalar, we have%
\begin{equation}%
{\displaystyle\int}
d^{D}xI^{\mu}\xi_{\mu}=0, \label{eqn7}%
\end{equation}
and so, both equations (\ref{eqn4}) and (\ref{eqn7}) are equivalent. Because
the DIs are constructed to be linear in the EDs, by performing integration(s)
by parts and/or redefining dummy indices in (\ref{eqn7}) we can easily obtain%

\begin{equation}%
{\displaystyle\int}
d^{D}x\left[  \underset{\delta e_{\mu\left(  \alpha\right)  }%
}{\underbrace{\left(  ...\right)  }}E^{\mu\left(  \alpha\right)
}+\underset{\delta\omega_{\mu\left(  \alpha\beta\right)  }%
}{\underbrace{\left(  ...\right)  }}E^{\mu\left(  \alpha\beta\right)
}\right]  =%
{\displaystyle\int}
d^{D}x\left(  E^{\mu\left(  \alpha\right)  }\delta e_{\mu\left(
\alpha\right)  }+E^{\mu\left(  \alpha\beta\right)  }\delta\omega_{\mu\left(
\alpha\beta\right)  }\right)  \label{eqn8}%
\end{equation}
and we can just read off the transformation of the fields that correspond to a
particular DI.

Construction of a DI is an iterative procedure and we will illustrate it by
first finding DI that generates a rotational (\textquotedblleft
Lorentz\textquotedblright) invariance which is well-known for N-bein gravity.
Such a DI arises quite naturally, because the simplest DI that one can think
of is to start from the derivative of the ED (\ref{eqn6}). Let us consider
$E_{,\tau}^{\tau\left(  \rho\sigma\right)  }:$%

\[
E_{,\tau}^{\tau\left(  \rho\sigma\right)  }=-eB^{\mu\left(  \alpha\right)
\tau\left(  \rho\right)  \nu\left(  \beta\right)  }e_{\mu\left(
\alpha\right)  ,\tau}\omega_{\nu~~\beta)}^{~(\sigma}+eB^{\mu\left(
\alpha\right)  \tau\left(  \sigma\right)  \nu\left(  \beta\right)  }%
e_{\mu\left(  \alpha\right)  ,\tau}\omega_{\nu~~\beta)}^{~(\rho}%
\]

\begin{equation}
-eA^{\tau\left(  \rho\right)  \nu\left(  \beta\right)  }\omega_{\nu
~~\beta),\tau}^{~(\sigma}+eA^{\tau\left(  \sigma\right)  \nu\left(
\beta\right)  }\omega_{\nu~~\beta),\tau}^{~(\rho}. \label{eqn9}%
\end{equation}
(The first two terms in $E_{,\tau}^{\tau\left(  \rho\sigma\right)  }$ are
$eC^{\lambda\left(  \gamma\right)  \nu\left(  \alpha\right)  \mu\left(
\rho\right)  \tau\left(  \sigma\right)  }e_{\lambda\left(  \gamma\right)
,\tau}e_{\nu\left(  \alpha\right)  ,\mu}+eB^{\nu\left(  \alpha\right)
\mu\left(  \rho\right)  \tau\left(  \sigma\right)  }e_{\nu\left(
\alpha\right)  ,\mu\tau}$ (using (\ref{eqn46a}) for $\left(  eB^{\nu\left(
\alpha\right)  \mu\left(  \rho\right)  \tau\left(  \sigma\right)  }\right)
,_{\tau}$), but both of them vanish because of antisymmetry of $C$ and $B$
(see Appendix A).)

We start the iterative procedure by trying to express all terms with
derivatives appearing in (\ref{eqn9}) in terms of EDs by comparison with the
corresponding contributions in (\ref{eqn5}) and (\ref{eqn6}). Two first terms
of (\ref{eqn9}) are exactly the first term in (\ref{eqn6}) contracted with a
connection. The last two terms are not exactly equal to any of the EDs, but
because of the presence of derivatives of connections, they can be related
only to (\ref{eqn5}). It is obvious that trying to relate them to (\ref{eqn5})
we have to perform a contraction of $B$ with the N-bein field to obtain an
expression similar to (\ref{eqn42a}). Because $E^{\tau\left(  \rho\right)  }$
has only one free external index $\tau$, then it must be a contraction of this
index and this fixes the choice of the indices in the expansion of $B$ needed
to simplify this calculation. Contraction of (\ref{eqn5}) with $e_{\tau
}^{\left(  \sigma\right)  }$ gives%

\begin{equation}
e_{\tau}^{\left(  \sigma\right)  }E^{\tau\left(  \rho\right)  }=-ee_{\tau
}^{\left(  \sigma\right)  }B^{\tau\left(  \rho\right)  \mu\left(
\alpha\right)  \nu\left(  \beta\right)  }\omega_{\nu\left(  \alpha
\beta\right)  ,\mu}-ee_{\tau}^{\left(  \sigma\right)  }B^{\tau\left(
\rho\right)  \mu\left(  \alpha\right)  \nu\left(  \beta\right)  }\omega
_{\mu\left(  \alpha\gamma\right)  }\omega_{\nu~~\beta)}^{~(\gamma}.
\label{eqn10a}%
\end{equation}
Using (\ref{eqn42a}) for the first term of the right-hand side of
(\ref{eqn10a}) and solving it for $eA^{\mu\left(  \rho\right)  \nu\left(
\beta\right)  }\omega_{\nu~~\beta),\mu}^{~(\sigma}$ we obtain%

\begin{equation}
eA^{\mu\left(  \rho\right)  \nu\left(  \beta\right)  }\omega_{\nu~~\beta),\mu
}^{~(\sigma}=\frac{1}{2}\left[  e_{\tau}^{\left(  \sigma\right)  }\left(
E^{\tau\left(  \rho\right)  }+eB^{\tau\left(  \rho\right)  \mu\left(
\alpha\right)  \nu\left(  \beta\right)  }\omega_{\mu\left(  \alpha
\gamma\right)  }\omega_{\nu~~\beta)}^{~(\gamma}\right)  +e\eta^{\left(
\sigma\right)  \left(  \rho\right)  }A^{\mu\left(  \alpha\right)  \nu\left(
\beta\right)  }\omega_{\nu\left(  \alpha\beta\right)  ,\mu}\right]  .
\label{eqn10}%
\end{equation}
Substituting the expression $eA^{\mu\left(  \rho\right)  \nu\left(
\beta\right)  }\omega_{\nu~~\beta),\mu}^{~(\sigma}$ from (\ref{eqn10}) into
the second line of (\ref{eqn9}) and using the identity%

\begin{equation}
eB^{\mu\left(  \alpha\right)  \tau\left(  \rho\right)  \nu\left(
\beta\right)  }e_{\mu\left(  \alpha\right)  ,\tau}=E^{\nu\left(  \rho
\beta\right)  }+eA^{\nu\left(  \rho\right)  \tau\left(  \lambda\right)
}\omega_{\tau~~\lambda)}^{~(\beta}-eA^{\nu\left(  \beta\right)  \tau\left(
\lambda\right)  }\omega_{\tau~~\lambda)}^{~(\rho} \label{eqn11}%
\end{equation}
in the first line of (\ref{eqn9}), gives us%

\begin{equation}
E_{,\tau}^{\tau\left(  \rho\sigma\right)  }=-E^{\nu\left(  \rho\beta\right)
}\omega_{\nu~~\beta)}^{~(\sigma}+E^{\nu\left(  \sigma\beta\right)  }%
\omega_{\nu~~\beta)}^{~(\rho}-\frac{1}{2}e_{\tau}^{\left(  \sigma\right)
}E^{\tau\left(  \rho\right)  }+\frac{1}{2}e_{\tau}^{\left(  \rho\right)
}E^{\tau\left(  \sigma\right)  }+W^{\left(  \rho\sigma\right)  } \label{eqn12}%
\end{equation}
where, as a result of the first iteration, all terms with derivatives are
expressed in terms of EDs. The remaining terms, all of them without
derivatives, are collected in $W^{\left(  \rho\sigma\right)  }:$%

\[
W^{\left(  \rho\sigma\right)  }=-\left(  eA^{\nu\left(  \rho\right)
\tau\left(  \lambda\right)  }\omega_{\tau~~\lambda)}^{~(\beta}-eA^{\nu\left(
\beta\right)  \tau\left(  \lambda\right)  }\omega_{\tau~~\lambda)}^{~(\rho
}\right)  \omega_{\nu~~\beta)}^{~(\sigma}+\left(  eA^{\nu\left(
\sigma\right)  \tau\left(  \lambda\right)  }\omega_{\tau~~\lambda)}^{~(\beta
}-eA^{\nu\left(  \beta\right)  \tau\left(  \lambda\right)  }\omega
_{\tau~~\lambda)}^{~(\sigma}\right)  \omega_{\nu~~\beta)}^{~(\rho}%
\]

\begin{equation}
-\frac{1}{2}e_{\tau}^{\left(  \sigma\right)  }\left[  eB^{\tau\left(
\rho\right)  \mu\left(  \alpha\right)  \nu\left(  \beta\right)  }\omega
_{\mu\left(  \alpha\gamma\right)  }\omega_{\nu~~\beta)}^{~(\gamma}\right]
+\frac{1}{2}e_{\tau}^{\left(  \rho\right)  }\left[  eB^{\tau\left(
\sigma\right)  \mu\left(  \alpha\right)  \nu\left(  \beta\right)  }\omega
_{\mu\left(  \alpha\gamma\right)  }\omega_{\nu~~\beta)}^{~(\gamma}\right]  .
\label{eqn13}%
\end{equation}
After performing an expansion of $B$ and contracting with $e_{\tau}^{\left(
\sigma\right)  }$ using (\ref{eqn42a}), all terms have the same structure:
$A\omega\omega.$ A relabelling the dummy indices leads to a complete
cancellation, so that $W^{\left(  \rho\sigma\right)  }=0$. The iterative
procedure now stops and the rotational DI is%

\begin{equation}
I^{\left(  \rho\sigma\right)  }=E_{,\nu}^{\nu\left(  \rho\sigma\right)
}+E^{\nu\left(  \rho\beta\right)  }\omega_{\nu~~\beta)}^{~(\sigma}%
-E^{\nu\left(  \sigma\beta\right)  }\omega_{\nu~~\beta)}^{~(\rho}-\frac{1}%
{2}e_{\nu}^{\left(  \rho\right)  }E^{\nu\left(  \sigma\right)  }+\frac{1}%
{2}e_{\nu}^{\left(  \sigma\right)  }E^{\nu\left(  \rho\right)  }=0.
\label{eqn14}%
\end{equation}
Upon substitution of (\ref{eqn14}), contracted with the parameter of
corresponding tensorial dimension and symmetries,\textit{ }$r_{\left(
\gamma\sigma\right)  }$\textit{, } into (\ref{eqn8}), making simple
rearrangements and integration by parts, we have%

\begin{equation}
dS=%
{\displaystyle\int}
d^{D}xI^{\left(  \gamma\sigma\right)  }r_{\left(  \gamma\sigma\right)  }=%
{\displaystyle\int}
d^{D}x\left[  \underset{\delta e_{\nu\left(  \sigma\right)  }%
}{\underbrace{\left(  -e_{\nu}^{\left(  \gamma\right)  }r_{\left(
\gamma\sigma\right)  }\right)  }}E^{\nu\left(  \sigma\right)  }%
+\underset{\delta\omega_{\nu\left(  \rho\sigma\right)  }}{\underbrace{\left(
-r_{\left(  \rho\sigma\right)  ,\nu}+\omega_{\nu~~\sigma)}^{~(\gamma
}r_{\left(  \rho\gamma\right)  }-\omega_{\nu~~\rho)}^{~(\gamma}r_{\left(
\sigma\gamma\right)  }\right)  }}E^{\nu\left(  \rho\sigma\right)  }\right]
\label{eqn15}%
\end{equation}
which gives the well-known rotational invariance for the first order N-bein
gravity (the same invariance was derived using the Hamiltonian approach to
N-bein gravity in \cite{3D, Report, Darboux})%

\begin{equation}
\delta_{r}e_{\nu\left(  \sigma\right)  }=-e_{\nu}^{\left(  \gamma\right)
}r_{\left(  \gamma\sigma\right)  }~, \label{eqn16}%
\end{equation}

\begin{equation}
\delta_{r}\omega_{\nu\left(  \rho\sigma\right)  }=-r_{\left(  \rho
\sigma\right)  ,\nu}+\omega_{\nu~~\sigma)}^{~(\gamma}r_{\left(  \rho
\gamma\right)  }-\omega_{\nu~~\rho)}^{~(\gamma}r_{\left(  \sigma\gamma\right)
}~. \label{eqn17}%
\end{equation}

Schwinger in \cite{Schwinger} started from the transformations (\ref{eqn16}%
)-(\ref{eqn17}) (see page 1254, unnumbered equation in the middle of the
second column) and from them obtained the identity (\ref{eqn14}). This is the
standard way of \textquotedblleft building\textquotedblright\ a DI by using a
\textit{known }invariance \cite{Noether}. This actually underestimates the
significance of the Lagrangian approach as DIs can be constructed from the
Euler derivatives and consequently the invariance can be found from the DIs.
Schwinger then used invariance of the fields $e_{\tau\left(  \sigma\right)  }$
and $\omega_{\tau\left(  \rho\sigma\right)  }$, which are both covariant
vectors in external indices, under the infinitesimal coordinate
transformations in the external space (diffeomorphism)%

\begin{equation}
\delta_{diff}e_{\tau\left(  \sigma\right)  }=-e_{\tau\left(  \sigma\right)
,\nu}\xi^{\nu}-e_{\nu\left(  \sigma\right)  }\xi_{,\tau}^{\nu}, \label{eqn18}%
\end{equation}

\begin{equation}
\delta_{diff}\omega_{\tau\left(  \rho\sigma\right)  }=-\omega_{\tau\left(
\rho\sigma\right)  ,\nu}\xi^{\nu}-\omega_{\nu\left(  \rho\sigma\right)  }%
\xi_{,\tau}^{\nu} \label{eqn19}%
\end{equation}
and obtained the second DI by substitution of (\ref{eqn18}) and (\ref{eqn19})
into (\ref{eqn8}) and then converting it into the form of (\ref{eqn7})%

\begin{equation}
I_{\nu}=-E^{\tau\left(  \sigma\right)  }e_{\tau\left(  \sigma\right)  ,\nu
}+\left(  E^{\tau\left(  \sigma\right)  }e_{\nu\left(  \sigma\right)
}\right)  _{,\tau}-E^{\tau\left(  \rho\sigma\right)  }\omega_{\tau\left(
\rho\sigma\right)  ,\nu}+\left(  E^{\tau\left(  \rho\sigma\right)  }%
\omega_{\nu\left(  \rho\sigma\right)  }\right)  _{,\tau}=0. \label{eqn20}%
\end{equation}
(This is Schwinger's equation in our notation.) We note that $\delta
_{diff}\left(  ..\right)  $ cannot be obtained from the first class
constraints of the Hamiltonian formulation of the EC action \cite{Report,
Darboux, affine-metric} even though it is the transformation that leaves the
action invariant.

Instead of (\ref{eqn20}), we will build the second DI in a way similar to the
rotational one, starting from derivatives of the second ED (\ref{eqn5}). If
such an identity exists, we have translational invariance in the tangent
space. By performing explicit differentiation of (\ref{eqn5}) and using
(\ref{eqn46a}) we obtain%

\begin{equation}
E_{,\tau}^{\tau\left(  \rho\right)  }=-eC^{\lambda\left(  \sigma\right)
\tau\left(  \rho\right)  \mu\left(  \alpha\right)  \nu\left(  \beta\right)
}e_{\lambda\left(  \sigma\right)  ,\tau}\left(  \omega_{\nu\left(  \alpha
\beta\right)  ,\mu}+\omega_{\mu\left(  \alpha\gamma\right)  }\omega
_{\nu~~\beta)}^{~(\gamma}\right)  -eB^{\tau\left(  \rho\right)  \mu\left(
\alpha\right)  \nu\left(  \beta\right)  }\left(  \omega_{\nu\left(
\alpha\beta\right)  ,\mu}+\omega_{\mu\left(  \alpha\gamma\right)  }\omega
_{\nu~~\beta)}^{~(\gamma}\right)  _{,\tau}. \label{eqn21}%
\end{equation}
In contrast to the rotational DI (\ref{eqn14}), here we have terms quadratic
in the derivatives $e_{\lambda\left(  \sigma\right)  ,\tau}$ and $\omega
_{\nu\left(  \alpha\beta\right)  ,\mu}$. Terms with either $e_{\lambda\left(
\sigma\right)  ,\tau}$ or $\omega_{\nu\left(  \alpha\beta\right)  ,\mu}$ are
in the EDs (\ref{eqn5}) and (\ref{eqn6}) when contracted with $B$. This
suggests an expansion of $C$ using the only one free (internal) index $\left(
\rho\right)  $ as in this case we have $B$'s with all indices which appear in
front of either $e_{\lambda\left(  \sigma\right)  ,\tau}$ or $\omega
_{\nu\left(  \alpha\beta\right)  ,\mu}$. When compared with the last step in
the derivation of the rotational DI (where we had a coefficient $A$ and it was
necessary to contract $B$ with $e_{\tau}^{\left(  \sigma\right)  }$), the ED
here is different: we have the coefficient $C$ which is the next
\textquotedblleft generation\textquotedblright\ of the $ABC$ functions and it
has to be expanded in term of $B$'s. Using (\ref{eqn46}) we obtain%

\[
E_{,\tau}^{\tau\left(  \rho\right)  }=-ee^{\tau\left(  \rho\right)
}B^{\lambda\left(  \sigma\right)  \mu\left(  \alpha\right)  \nu\left(
\beta\right)  }\left(  \omega_{\nu\left(  \alpha\beta\right)  ,\mu}%
+\omega_{\mu\left(  \alpha\gamma\right)  }\omega_{\nu~~\beta)}^{~(\gamma
}\right)  e_{\lambda\left(  \sigma\right)  ,\tau}%
\]

\[
+ee^{\lambda\left(  \rho\right)  }B^{\mu\left(  \sigma\right)  \nu\left(
\alpha\right)  \tau\left(  \beta\right)  }\left(  \omega_{\nu\left(
\alpha\beta\right)  ,\mu}+\omega_{\mu\left(  \alpha\gamma\right)  }\omega
_{\nu~~\beta)}^{~(\gamma}\right)  e_{\lambda\left(  \sigma\right)  ,\tau}%
\]

\[
-ee^{\mu\left(  \rho\right)  }B^{\nu\left(  \sigma\right)  \tau\left(
\alpha\right)  \lambda\left(  \beta\right)  }e_{\lambda\left(  \sigma\right)
,\tau}\left(  \omega_{\nu\left(  \alpha\beta\right)  ,\mu}+\omega_{\mu\left(
\alpha\gamma\right)  }\omega_{\nu~~\beta)}^{~(\gamma}\right)
\]

\[
+ee^{\nu\left(  \rho\right)  }B^{\tau\left(  \sigma\right)  \lambda\left(
\alpha\right)  \mu\left(  \beta\right)  }e_{\lambda\left(  \sigma\right)
,\tau}\left(  \omega_{\nu\left(  \alpha\beta\right)  ,\mu}+\omega_{\mu\left(
\alpha\gamma\right)  }\omega_{\nu~~\beta)}^{~(\gamma}\right)
\]

\begin{equation}
-eB^{\tau\left(  \rho\right)  \mu\left(  \alpha\right)  \nu\left(
\beta\right)  }\left(  \omega_{\mu\left(  \alpha\gamma\right)  }\omega
_{\nu~~\beta)}^{~(\gamma}\right)  _{,\tau}. \label{eqn22}%
\end{equation}
(Here we used the identity (\ref{eqn50c}): $eB^{\tau\left(  \rho\right)
\mu\left(  \alpha\right)  \nu\left(  \beta\right)  }\omega_{\nu\left(
\alpha\beta\right)  ,\mu\tau}=0$.)

In the first two lines, we immediately have a full expression for the
contracted ED (\ref{eqn5}) and in the third and fourth lines we have
contributions which are equal to terms with derivatives in the contracted ED
(\ref{eqn6}). As the result, we convert (\ref{eqn22}) into%

\[
E_{,\tau}^{\tau\left(  \rho\right)  }=e^{\tau\left(  \rho\right)  }%
E^{\lambda\left(  \sigma\right)  }e_{\lambda\left(  \sigma\right)  ,\tau
}-e^{\lambda\left(  \rho\right)  }E^{\tau\left(  \sigma\right)  }%
e_{\lambda\left(  \sigma\right)  ,\tau}%
\]

\[
+e^{\mu\left(  \rho\right)  }E^{\nu\left(  \alpha\beta\right)  }\left(
\omega_{\nu\left(  \alpha\beta\right)  ,\mu}+\omega_{\mu\left(  \alpha
\gamma\right)  }\omega_{\nu~~\beta)}^{~(\gamma}\right)  -e^{\nu\left(
\rho\right)  }E^{\mu\left(  \alpha\beta\right)  }\left(  \omega_{\nu\left(
\alpha\beta\right)  ,\mu}+\omega_{\mu\left(  \alpha\gamma\right)  }\omega
_{\nu~~\beta)}^{~(\gamma}\right)  +W^{\left(  \rho\right)  }%
\]
where%

\[
W^{\left(  \rho\right)  }=e^{\mu\left(  \rho\right)  }\left[  eA^{\nu\left(
\alpha\right)  \nu^{\prime}\left(  \beta^{\prime}\right)  }\omega_{\nu
^{\prime}~~\beta^{\prime})}^{~(\beta}-eA^{\nu\left(  \beta\right)  \nu
^{\prime}\left(  \beta^{\prime}\right)  }\omega_{\nu^{\prime}~~\beta^{\prime
})}^{~(\alpha}\right]  \left(  \omega_{\nu\left(  \alpha\beta\right)  ,\mu
}+\omega_{\mu\left(  \alpha\gamma\right)  }\omega_{\nu~~\beta)}^{~(\gamma
}\right)
\]

\[
-e^{\nu\left(  \rho\right)  }\left[  eA^{\mu\left(  \alpha\right)  \nu
^{\prime}\left(  \beta^{\prime}\right)  }\omega_{\nu^{\prime}~~\beta^{\prime
})}^{~(\beta}-eA^{\mu\left(  \beta\right)  \nu^{\prime}\left(  \beta^{\prime
}\right)  }\omega_{\nu^{\prime}~~\beta^{\prime})}^{~(\alpha}\right]  \left(
\omega_{\nu\left(  \alpha\beta\right)  ,\mu}+\omega_{\mu\left(  \alpha
\gamma\right)  }\omega_{\nu~~\beta)}^{~(\gamma}\right)
\]

\begin{equation}
-eB^{\tau\left(  \rho\right)  \mu\left(  \alpha\right)  \nu\left(
\beta\right)  }\left(  \omega_{\mu\left(  \alpha\gamma\right)  ,\tau}%
\omega_{\nu~~\beta)}^{~(\gamma}+\omega_{\mu\left(  \alpha\gamma\right)
}\omega_{\nu~~\beta),\tau}^{~(\gamma}\right)  . \label{eqn23}%
\end{equation}

All terms with two derivatives have been expressed in terms of EDs from
(\ref{eqn5}) and (\ref{eqn6}) and we are left only with terms linear in the
derivatives and terms without derivatives which are of third order in the
connections. We have to continue and try to express terms with derivatives
again through EDs.

The derivatives presented in $W^{\left(  \rho\right)  }$ correspond to the ED
(\ref{eqn5}), $E^{\lambda\left(  \sigma\right)  }$, contracted with
$\omega_{\mu\left(  \alpha\gamma\right)  }.$ Since $e^{\mu\left(  \rho\right)
}A^{\nu\left(  \alpha\right)  \nu^{\prime}\left(  \beta^{\prime}\right)  }$
has the dimension of $B$, we can try to expand $B$ to obtain such
combinations. However, because $A$ (see (\ref{eqn23})) has only two external
indices that are contracted with indices of derivative of connections, it is
not immediately in the form of (\ref{eqn5}). Performing an expansion of
$B^{\tau\left(  \rho\right)  \mu\left(  \alpha\right)  \nu\left(
\beta\right)  }$ in a free internal index, (\ref{eqn45}), we obtain%

\begin{equation}
e^{\mu\left(  \rho\right)  }A^{\nu\left(  \alpha\right)  \tau\left(
\beta\right)  }-e^{\nu\left(  \rho\right)  }A^{\mu\left(  \alpha\right)
\tau\left(  \beta\right)  }=-e^{\tau\left(  \rho\right)  }A^{\mu\left(
\alpha\right)  \nu\left(  \beta\right)  }-B^{\tau\left(  \rho\right)
\mu\left(  \alpha\right)  \nu\left(  \beta\right)  }. \label{eqn25}%
\end{equation}

This is exactly the combination that we have in (\ref{eqn23}) and this allows
us to convert (\ref{eqn23}) into a contraction of the expression that we used
in our calculations of the rotational DI (\ref{eqn10}). In addition, after
substitution of (\ref{eqn25}) the last term in (\ref{eqn23}) cancels. At this
step, all derivatives have been eliminated and the rest of the terms, all of
third order in the connections, can be easily analyzed. As in the rotational
case, all of them cancel, and we have a new, translational, DI%

\begin{equation}
I^{\left(  \rho\right)  }=E_{,\nu}^{\nu\left(  \rho\right)  }-e^{\tau\left(
\rho\right)  }E^{\nu\left(  \beta\right)  }e_{\nu\left(  \beta\right)  ,\tau
}+e^{\tau\left(  \rho\right)  }E^{\nu\left(  \beta\right)  }e_{\tau\left(
\beta\right)  ,\nu}-e^{\mu\left(  \rho\right)  }E^{\nu\left(  \alpha
\beta\right)  }R_{\mu\nu\left(  \alpha\beta\right)  }+e^{\tau\left(
\rho\right)  }\omega_{\tau(\gamma\beta)}^{~}e_{\nu}^{\left(  \gamma\right)
}E^{\nu\left(  \beta\right)  }=0. \label{eqn26}%
\end{equation}

Contracting this with the corresponding parameter $t_{\left(  \rho\right)  }$
(in a way similar to what was done for the rotational DI in (\ref{eqn15}%
)),\textit{ }we obtain%

\begin{equation}
\int d^{D}x~I^{\left(  \rho\right)  }t_{\left(  \rho\right)  }=0.
\label{eqn27}%
\end{equation}
A simple rearrangement gives the transformations corresponding to this DI%

\begin{equation}
\delta_{t}e_{\nu\left(  \beta\right)  }=-t_{\left(  \beta\right)  ,\nu
}-e^{\tau\left(  \rho\right)  }e_{\nu\left(  \beta\right)  ,\tau}t_{\left(
\rho\right)  }+e^{\tau\left(  \rho\right)  }e_{\tau\left(  \beta\right)  ,\nu
}t_{\left(  \rho\right)  }+e^{\tau\left(  \rho\right)  }\omega_{\tau
(\gamma\beta)}^{~}e_{\nu}^{\left(  \gamma\right)  }t_{\left(  \rho\right)  },
\label{eqn28a}%
\end{equation}

\begin{equation}
\delta_{t}\omega_{\nu\left(  \alpha\beta\right)  }=-e^{\mu\left(  \rho\right)
}R_{\mu\nu\left(  \alpha\beta\right)  }t_{\left(  \rho\right)  }.
\label{eqn29}%
\end{equation}

Note that the identity (\ref{eqn26}) can be rewritten in a slightly different
form. If we add and subtract the term $\omega_{\nu(\beta}^{~~~~\rho)}%
E^{\nu\left(  \beta\right)  }$ and rewrite one of these terms as
$e^{\tau\left(  \rho\right)  }e_{\tau}^{\left(  \gamma\right)  }\omega
_{\nu\left(  \beta\gamma\right)  }E^{\nu\left(  \beta\right)  }$, then
$I^{\left(  \rho\right)  }$ becomes%

\begin{equation}
I^{\left(  \rho\right)  }=E_{,\nu}^{\nu\left(  \rho\right)  }-\omega
_{\nu(\beta}^{~~~~\rho)}E^{\nu\left(  \beta\right)  }+e^{\tau\left(
\rho\right)  }T_{\nu\tau\left(  \beta\right)  }E^{\nu\left(  \beta\right)
}-e^{\mu\left(  \rho\right)  }E^{\nu\left(  \alpha\beta\right)  }R_{\mu
\nu\left(  \alpha\beta\right)  } \label{eqn26a}%
\end{equation}
where%

\[
T_{\nu\tau\left(  \beta\right)  }\equiv e_{\tau\left(  \beta\right)  ,\nu
}-e_{\nu\left(  \beta\right)  ,\tau}+\omega_{\tau\left(  \gamma\beta\right)
}e_{\nu}^{\left(  \gamma\right)  }-\omega_{\nu\left(  \gamma\beta\right)
}e_{\tau}^{\left(  \gamma\right)  }%
\]
is the torsion tensor (see, e.g. \cite{DZ}),\footnote{Note that $T_{\nu
\tau\left(  \beta\right)  }$ is different from the torsion tensor
$\Gamma_{\left[  \alpha\beta\right]  }^{\gamma}$ which appears in metric
gravity when the affine connection $\Gamma_{\alpha\beta}^{\gamma}$ is assumed
to be nonsymmetric: $\Gamma_{\left[  \alpha\beta\right]  }^{\gamma}=\frac
{1}{2}\left(  \Gamma_{\alpha\beta}^{\gamma}-\Gamma_{\beta\alpha}^{\gamma
}\right)  $ (see, e.g. \cite{Carmeli}).} which is antisymmetric in external
indices $\nu$ and $\tau$. With $I^{\left(  \rho\right)  }$ of (\ref{eqn26a}),
the transformation $\delta_{t}e_{\nu\left(  \beta\right)  }$ takes the form%

\begin{equation}
\delta_{t}e_{\nu\left(  \beta\right)  }=-t_{\left(  \beta\right)  ,\nu}%
-\omega_{\nu(\beta}^{~~~~\rho)}t_{\left(  \rho\right)  }+e^{\tau\left(
\rho\right)  }T_{\nu\tau\left(  \beta\right)  }t_{\left(  \rho\right)  }.
\label{eqn28}%
\end{equation}

The first order formulation with two sets of independent variables leads to
two sets of EDs, two the simplest, basic, DIs and two symmetries. Many other
DIs can be constructed from these two basic identities and the corresponding
symmetries of an action can be found, and one particular DI (out of many
possible) gives the diffeomorphism%

\begin{equation}
I_{\nu}=e_{\nu\left(  \alpha\right)  }I^{\left(  \alpha\right)  }+\omega
_{\nu\left(  \gamma\beta\right)  }I^{\left(  \gamma\beta\right)  },
\label{eqn-diff}%
\end{equation}
which coincides with (\ref{eqn20}).

\section{Group properties of the transformations}

The algebraic properties of the transformations found in the previous Section
can be analyzed by considering commutators of two successive transformations.
(This is what Bergmann and Komar \cite{Bergmann} used to discuss group
properties of diffeomorphism invariance of the Einstein-Hilbert action where
the basic DI indeed is the \textquotedblleft world\textquotedblright\ vector,
contrary to the basic DIs of the Einstein-Cartan action.)

We find the commutators between two transformations of fields and using this
identify what algebra they obey. Let us start by calculating the commutator
between two rotational transformations $\left(  \delta_{r^{\prime\prime}%
}\delta_{r^{\prime}}-\delta_{r^{\prime}}\delta_{r^{\prime\prime}}\right)
e_{\tau\left(  \sigma\right)  }$. Using (\ref{eqn16}) we obtain%

\begin{equation}
\left(  \delta_{r^{\prime\prime}}\delta_{r^{\prime}}-\delta_{r^{\prime}}%
\delta_{r^{\prime\prime}}\right)  e_{\tau\left(  \sigma\right)  }=-e_{\tau
}^{\left(  \lambda\right)  }\bar{r}_{(\lambda\sigma)}=\delta_{\bar{r}}%
e_{\tau\left(  \sigma\right)  } \label{ct1}%
\end{equation}
where the new parameter is related to $r^{\prime}$ and $r^{\prime\prime}$ as%

\begin{equation}
\bar{r}_{(\lambda\sigma)}\equiv r_{(\lambda}^{\prime~\gamma)}r_{(\gamma
\sigma)}^{\prime\prime}-r_{(\lambda}^{\prime\prime~\gamma)}r_{(\gamma\sigma
)}^{\prime}. \label{ct2}%
\end{equation}

Similarly, using (\ref{eqn17}) we can find the commutator between two
rotations for a connection%

\begin{equation}
\left(  \delta_{r^{\prime\prime}}\delta_{r^{\prime}}-\delta_{r^{\prime}}%
\delta_{r^{\prime\prime}}\right)  \omega_{\tau\left(  \rho\sigma\right)
}=-\bar{r}_{\left(  \rho\sigma\right)  ,\tau}+\omega_{\tau~~\sigma
)}^{~(\lambda}\bar{r}_{\left(  \rho\lambda\right)  }-\omega_{\tau~~\rho
)}^{~(\lambda}\bar{r}_{\left(  \sigma\lambda\right)  }=\delta_{\bar{r}}%
\omega_{\tau\left(  \rho\sigma\right)  } \label{ct3}%
\end{equation}
with the same parameter $\bar{r}_{(\lambda\sigma)}$ as in (\ref{ct2}).

For the commutator between a rotation and a translation we obtain%

\begin{align}
\left(  \delta_{r}\delta_{t}-\delta_{t}\delta_{r}\right)  e_{\nu\left(
\beta\right)  }  &  =-\left(  r_{~~\beta)}^{(\rho}t_{\left(  \rho\right)
}\right)  _{,\nu}+e^{\lambda\left(  \gamma\right)  }e_{\lambda\left(
\beta\right)  ,\nu}\left(  r_{~~\gamma)}^{(\rho}t_{\left(  \rho\right)
}\right)  -e^{\lambda\left(  \gamma\right)  }e_{\nu\left(  \beta\right)
,\lambda}\left(  r_{~~\gamma)}^{(\rho}t_{\left(  \rho\right)  }\right)
\label{ct4}\\
&  +e^{\tau\left(  \sigma\right)  }\omega_{\tau\left(  \gamma\beta\right)
}e_{\nu}^{\left(  \gamma\right)  }\left(  r_{~~\sigma)}^{(\rho}t_{\left(
\rho\right)  }\right)  =\delta_{\bar{t}}e_{\nu\left(  \beta\right)
},\nonumber
\end{align}

\begin{equation}
\left(  \delta_{r}\delta_{t}-\delta_{t}\delta_{r}\right)  \omega_{\tau\left(
\rho\sigma\right)  }=-e^{\mu\left(  \beta\right)  }R_{\mu\tau\left(
\rho\sigma\right)  }\left(  r_{~~\beta)}^{(\gamma}t_{\left(  \gamma\right)
}\right)  =\delta_{\bar{t}}\omega_{\tau\left(  \rho\sigma\right)  }
\label{ct6}%
\end{equation}
with the new parameter%

\begin{equation}
\bar{t}_{\left(  \beta\right)  }\equiv r_{~~\beta)}^{(\rho}t_{\left(
\rho\right)  }. \label{ct7}%
\end{equation}

Calculation of the commutator between two translations is more involved. For
$e_{\nu\left(  \beta\right)  }$ it gives%

\begin{align}
\left(  \delta_{t^{\prime\prime}}\delta_{t^{\prime}}-\delta_{t^{\prime}}%
\delta_{t^{\prime\prime}}\right)  e_{\nu\left(  \beta\right)  }  &
=-e^{\lambda\left(  \rho\right)  }e^{\tau\left(  \gamma\right)  }%
R_{\lambda\tau\left(  \alpha\beta\right)  }t_{\left(  \rho\right)  }^{\prime
}t_{\left(  \gamma\right)  }^{\prime\prime}e_{\nu}^{\left(  \alpha\right)
}-\left(  e^{\lambda\left(  \rho\right)  }e^{\tau\left(  \gamma\right)
}T_{\lambda\tau\left(  \beta\right)  }t_{\left(  \rho\right)  }^{\prime
}t_{\left(  \gamma\right)  }^{\prime\prime}\right)  _{,\nu}\label{ct7a}\\
&  -\omega_{\nu(\beta}^{~~~~\sigma)}e^{\lambda\left(  \rho\right)  }%
e^{\tau\left(  \gamma\right)  }T_{\lambda\tau\left(  \sigma\right)
}t_{\left(  \rho\right)  }^{\prime}t_{\left(  \gamma\right)  }^{\prime\prime
}+e^{\mu\left(  \sigma\right)  }T_{\nu\mu\left(  \beta\right)  }%
e^{\lambda\left(  \rho\right)  }e^{\tau\left(  \gamma\right)  }T_{\lambda
\tau\left(  \sigma\right)  }t_{\left(  \rho\right)  }^{\prime}t_{\left(
\gamma\right)  }^{\prime\prime}\nonumber\\
&  =\delta_{\tilde{r}}e_{\nu\left(  \beta\right)  }+\delta_{\tilde{t}}%
e_{\nu\left(  \beta\right)  }\nonumber
\end{align}
and for a connection $\omega_{\nu\left(  \alpha\beta\right)  }$ we obtain%

\begin{align}
\left(  \delta_{t^{\prime\prime}}\delta_{t^{\prime}}-\delta_{t^{\prime}}%
\delta_{t^{\prime\prime}}\right)  \omega_{\nu\left(  \alpha\beta\right)  }  &
=-\left(  e^{\lambda\left(  \rho\right)  }e^{\tau\left(  \gamma\right)
}R_{\lambda\tau\left(  \alpha\beta\right)  }t_{\left(  \rho\right)  }^{\prime
}t_{\left(  \gamma\right)  }^{\prime\prime}\right)  _{,\nu}\label{ct8}\\
&  +e^{\lambda\left(  \rho\right)  }e^{\tau\left(  \gamma\right)  }t_{\left(
\rho\right)  }^{\prime}t_{\left(  \gamma\right)  }^{\prime\prime}\left(
R_{\lambda\tau\left(  \sigma\beta\right)  }\omega_{\nu~~\alpha)}^{~(\sigma
}-R_{\lambda\tau\left(  \sigma\alpha\right)  }\omega_{\nu~~\beta)}^{~(\sigma
}\right) \nonumber\\
&  -e^{\lambda\left(  \rho\right)  }e^{\tau\left(  \gamma\right)  }t_{\left(
\rho\right)  }^{\prime}t_{\left(  \gamma\right)  }^{\prime\prime}%
T_{\lambda\tau\left(  \sigma\right)  }e^{\mu\left(  \sigma\right)  }R_{\mu
\nu\left(  \alpha\beta\right)  }\nonumber\\
&  =-\tilde{r}_{\left(  \alpha\beta\right)  ,\nu}+\omega_{\nu~~\beta
)}^{~(\gamma}\tilde{r}_{\left(  \alpha\gamma\right)  }-\omega_{\nu~~\alpha
)}^{~(\gamma}\tilde{r}_{\left(  \beta\gamma\right)  }-e^{\mu\left(
\sigma\right)  }R_{\mu\nu\left(  \alpha\beta\right)  }\tilde{t}_{\left(
\sigma\right)  }\nonumber\\
&  =\delta_{\tilde{r}}\omega_{\nu\left(  \alpha\beta\right)  }+\delta
_{\tilde{t}}\omega_{\nu\left(  \alpha\beta\right)  }\nonumber
\end{align}
where the new parameters are%

\begin{equation}
\tilde{r}_{\left(  \alpha\beta\right)  }\equiv e^{\lambda\left(  \rho\right)
}e^{\tau\left(  \gamma\right)  }R_{\lambda\tau\left(  \alpha\beta\right)
}t_{\left(  \rho\right)  }^{\prime}t_{\left(  \gamma\right)  }^{\prime\prime}
\label{ct9}%
\end{equation}
and%

\begin{equation}
\tilde{t}_{\left(  \sigma\right)  }\equiv e^{\lambda\left(  \rho\right)
}e^{\tau\left(  \gamma\right)  }T_{\lambda\tau\left(  \sigma\right)
}t_{\left(  \rho\right)  }^{\prime}t_{\left(  \gamma\right)  }^{\prime\prime}.
\label{ct10}%
\end{equation}

Note that in the flat spacetime limit both $R_{\lambda\tau\left(  \alpha
\beta\right)  }$ and $T_{\lambda\tau\left(  \sigma\right)  }$ are zero and in
this limit the ordinary Poincar\'{e} symmetry is recovered. In addition, it is
clear from (\ref{ct7a}) and (\ref{ct8}) that the algebra of translations is
not closed by itself (without rotation). In the second order formulation of
the EC action only one ID, $I^{\left(  \rho\right)  }$, can be found from the
variation of the Lagrangian with respect to $e_{\tau\left(  \rho\right)  }$
and the necessity of finding the second DI, $I^{\left(  \alpha\beta\right)  }%
$, is not obvious without having some \textit{a priori }knowledge. Possibly,
non-closure of the algebra of translations in the second order formulation
provides a clue that we should seek another DI.

Commutators (\ref{ct1})/(\ref{ct3}), (\ref{ct4})/(\ref{ct6}) and
(\ref{ct7a})/(\ref{ct8}) form a closed algebra; (\ref{ct1})/(\ref{ct3}) and
(\ref{ct4})/(\ref{ct6}) are part of the flat spacetime Poincar\'{e} algebra
while the commutators (\ref{ct7a})/(\ref{ct8}) differs from Poincar\'{e} and
similar to found in \cite{Trautman, Hehl}. Consequently, transformations that
follow from the basic identities form the group with field dependent structure
functions. These transformations are not the only ones that leave the EC
action invariant, as many other DIs can be constructed that correspond to
invariances of the action, and their group properties can be studied. For
example, the identity (\ref{eqn20})\textit{ }corresponds to diffeomorphism
invariance of (\ref{eqn18}) and (\ref{eqn19}) and related to rotation and
translation by (\ref{eqn-diff}). The commutator of two diffeomorphism
transformations gives%

\begin{equation}
\left(  \delta_{diff}^{\prime}\delta_{diff}^{\prime\prime}-\delta
_{diff}^{\prime\prime}\delta_{diff}^{\prime}\right)  e_{\nu\left(
\lambda\right)  }=-e_{\rho\left(  \lambda\right)  }\bar{\xi}_{,\nu}^{\rho
}-e_{\nu\left(  \lambda\right)  ,\rho}\bar{\xi}^{\rho}=\delta_{diff}%
e_{\nu\left(  \lambda\right)  } \label{dif4}%
\end{equation}
with the new parameter%

\begin{equation}
\bar{\xi}^{\rho}\equiv\xi_{,\gamma}^{\prime\rho}\xi^{\prime\prime\gamma}%
-\xi_{,\gamma}^{\prime\prime\rho}\xi^{\prime\gamma}\label{dif5}%
\end{equation}
which is the same as found by Bergmann and Komar in \cite{Bergmann}.\textit{ }

We can also calculate the commutators among transformations that correspond to
different symmetries, e.g. diffeomorphism and rotation%

\begin{align}
\left(  \delta_{r}\delta_{diff}-\delta_{diff}\delta_{r}\right)  e_{\nu\left(
\lambda\right)  }  &  =\delta_{r}\left(  -e_{\rho\left(  \lambda\right)  }%
\xi_{,\nu}^{\rho}-e_{\nu\left(  \lambda\right)  ,\rho}\xi^{\rho}\right)
-\delta_{diff}\left(  -e_{\tau}^{\left(  \gamma\right)  }r_{\left(
\gamma\sigma\right)  }\right) \label{dif2}\\
&  =e_{\rho}^{\left(  \gamma\right)  }r_{\left(  \gamma\lambda\right)  }%
\xi_{,\nu}^{\rho}-\left(  -e_{\tau}^{\left(  \gamma\right)  }r_{\left(
\gamma\sigma\right)  }\right)  _{,\rho}\xi^{\rho}-\left(  e_{\sigma}^{\left(
\gamma\right)  }\xi_{,\nu}^{\sigma}-e_{\nu,\sigma}^{\left(  \gamma\right)
}\xi^{\sigma}\right)  r_{\left(  \gamma\lambda\right)  }\nonumber\\
&  =e_{\nu}^{\left(  \gamma\right)  }r_{\left(  \gamma\lambda\right)  ,\rho
}\xi^{\rho}=-e_{\nu}^{\left(  \gamma\right)  }\hat{r}_{\left(  \gamma
\lambda\right)  }=\delta_{\hat{r}}e_{\nu}^{\left(  \gamma\right)  }\nonumber
\end{align}
where%

\begin{equation}
\hat{r}_{\left(  \gamma\lambda\right)  }\equiv r_{\left(  \gamma
\lambda\right)  ,\rho}\xi^{\rho}, \label{dif3}%
\end{equation}
which shows the group property (it gives transformation under rotation
again)\textit{.} Note that this is not equivalent with what is reported in
\cite{CNP, Henneaux, Maluf} where it is claimed that the first class
constraints that correspond to these transformations have zero PB. We found
the only one paper where the PB between the generator of rotation
(\textquotedblleft Lorentz\textquotedblright\ transformation) and the
generator of \textquotedblleft three-dimensional general coordinate
transformations\textquotedblright\ is proportional to the generator of
rotation \cite{Stefano}.

Finally we can consider a commutator of diffeomorphism and translational
transformations. This leads to the following result%

\begin{equation}
\left(  \delta_{t}\delta_{diff}-\delta_{diff}\delta_{t}\right)  e_{\nu\left(
\lambda\right)  }=\left(  t_{\left(  \lambda\right)  ,\rho}\xi^{\rho}\right)
_{,\nu}. \label{dif10}%
\end{equation}
It is clear that these two transformations do not form a group. This is
consistent with the fact that at the Lagrangian level diffeomorphism and
translation are not independent as DI of diffeomorphism can be expressed as a
linear combination of rotational and translational DIs (see (\ref{eqn-diff})).
At the Hamiltonian level, this probably is an indication that there is no
canonical transformation between variables in which the corresponding
Hamiltonian formulations of the same theory produce diffeomorphism and
translation in the internal space as gauge symmetries, respectively.

By construction of the DIs, the EC action (\ref{eqn1})-(\ref{eqn2}) is
automatically invariant \textquotedblleft off-shell\textquotedblright\ under
the transformations (\ref{eqn16})-(\ref{eqn17}) and (\ref{eqn28a}%
)-(\ref{eqn29}). Direct demonstration of this is most easily done if the $ABC$
properties given in Appendix A are used. It is not difficult to check that the
variation of the Lagrange density under rotation is $\delta_{r}L=0$ and under
translation is $\delta_{t}L=-\left(  ee^{\lambda\left(  \rho\right)
}Rt_{\left(  \rho\right)  }\right)  _{,\lambda}$ where $R=e^{\mu\left(
\alpha\right)  }e^{\nu\left(  \beta\right)  }R_{\mu\nu\left(  \alpha
\beta\right)  }$.

In addition, because $L\left(  e,\omega\right)  $ is equivalent to the second
order formulation $L\left(  e\right)  $, where $\omega_{\nu\left(  \alpha
\beta\right)  }$ is expressed in terms of $e_{\nu\left(  \beta\right)  }$ by
using its equation of motion and is not an independent field, it is
unnecessary to recalculate the DI to find transformations for $L\left(
e\right)  $. The solution for connections in terms of N-beins is known and can
be presented in terms of $A^{\gamma\left(  \tau\right)  \mu\left(
\varepsilon\right)  }$ (see \cite{Report})%

\begin{equation}
\omega_{\nu}^{~~\left(  \tau\lambda\right)  }=\frac{1}{2}e_{\nu\left(
\varepsilon\right)  }\left(  A^{\gamma\left(  \tau\right)  \mu\left(
\varepsilon\right)  }e_{\gamma,\mu}^{\left(  \lambda\right)  }+A^{\gamma
\left(  \varepsilon\right)  \mu\left(  \lambda\right)  }e_{\gamma,\mu
}^{\left(  \tau\right)  }-A^{\gamma\left(  \lambda\right)  \mu\left(
\tau\right)  }e_{\gamma,\mu}^{\left(  \varepsilon\right)  }\right)  .
\label{eqn30}%
\end{equation}
Upon substitution of (\ref{eqn30}) into (\ref{eqn28}) we obtain%

\begin{equation}
\delta_{t}e_{\nu\left(  \beta\right)  }=-t_{\left(  \beta\right)  ,\nu}%
-\omega_{\nu(\beta}^{~~~~\rho)}t_{\left(  \rho\right)  }. \label{eqn31}%
\end{equation}
The transformation of $e_{\nu\left(  \beta\right)  }$ under a rotation does
not change when passing to the second order formulation because (\ref{eqn16})
does not depend on the connection. The second transformation, (\ref{eqn29}),
for the second order formulation can be derived from (\ref{eqn30}) and
(\ref{eqn31}) if we express $R_{\mu\nu\left(  \alpha\beta\right)  }$ in terms
of $e_{\nu\left(  \beta\right)  }$ and its derivatives. All term proportional
to $T_{\nu\tau\left(  \beta\right)  }$ disappear in (\ref{eqn31}) if we move
to the second order formalism.

Note that the connection $\omega_{\nu(\beta}^{~~~~\rho)}$ in (\ref{eqn31}) is
a function of $e_{\nu\left(  \beta\right)  }$ given in (\ref{eqn30}) and is
not an independent variable, if we consider the second order formalism.
Ironically, exactly this equation, (\ref{eqn31}), has been used for decades as
the main argument\ that the EC action in not invariant under a translation in
dimensions higher than three. In the literature, equation (\ref{eqn31}) is
often accompanied by $\delta_{t}\omega_{\nu\left(  \alpha\beta\right)  }=0$
rather than (\ref{eqn29}) \cite{Grignani} (though in three dimensions it is
indeed true that $\delta_{t}\omega_{\nu\left(  \alpha\beta\right)  }=0$).

The transformations (\ref{eqn16}), (\ref{eqn17}) and (\ref{eqn28}),
(\ref{eqn29}) are not new and have been presented, e.g. in \cite{Hehl,
Leclerc}. However, the question whether these transformations represent the
gauge symmetry which follows from the Hamiltonian analysis was not discussed.

We consider basic IDs, (\ref{eqn14}) and (\ref{eqn26}), and one well known but
not basic, (\ref{eqn20}), which gives diffeomorphism transformation, but many
other DIs and symmetries can be found and group properties of the
corresponding transformations can be studied. All of them, as in the simple
example above (see (\ref{dif10})), cannot be combined into a group but
different groups can be formed (e.g. diffeomorphism and rotation form a group
as do translation and rotation), but not every group follows from the first
class constraints. The Hamiltonian analysis leads uniquely to the gauge
invariance (what follows from the structure of first class constraints)
without any ambiguity. Preliminary results of the Hamiltonian formulation of
the EC action \cite{3D, Report} show that even from first simple steps of the
Dirac procedure, from the tensorial dimension of primary first class
constraints, the translation and rotation in the tangent space follow
unambiguously as a gauge symmetry. In conclusion we argue that the Lagrangian
methods also allow us to single out a unique gauge invariance of any theory,
the same invariance as follows from the Hamiltonian analysis.

\section{Conclusion}

What invariances are true gauge invariances of the N-bein formulation of
gravity? \textquotedblleft What is a gauge symmetry and what is
not?\textquotedblright\ \cite{Matschull}. The rotational and translational
transformations, (\ref{eqn16}), (\ref{eqn17}) and (\ref{eqn28}),
(\ref{eqn29}), both have internal parameters and if a rotational invariance is
a gauge invariance, then should translational invariance. In the Hamiltonian
analysis of the first order formulation of N-bein gravity, exactly these two
parameters emerge, both with only internal indices \cite{Report, 3D, Darboux}.
This result which contradicts conventional wisdom\ motivated us to perform the
short investigation reported here, in which we examine what invariances appear
in the Lagrangian approach. In the Hamiltonian formulation, where elimination
of the second class constraints is needed \cite{Diracbook}, restoration of the
manifestly covariant form of the transformations is difficult. If one is not
interested in finding the constraints and their algebra, the manifestly
covariant Lagrangian method is simpler. However, there is an apparent
\textquotedblleft deficiency\textquotedblright\ in the Lagrangian method: we
presented here a few differential identities (see (\ref{eqn14}),
(\ref{eqn20}), (\ref{eqn26})), but linear combinations of these identities
will lead to additional invariances of an original action. These cannot all be
gauge symmetries generated by the first class constraints. This is clear, as
in the Hamiltonian formulation each gauge symmetry originates from primary
first class constraints and, if all symmetries are to be accommodated, more
first class constraints than variables would be needed. This would be
inconsistent as it leads to a negative number of degrees of freedom which is
an unacceptable result. Similarly, both translational and diffeomorphism
invariances cannot be simultaneously gauge symmetries arising from the first
class constraints in any Hamiltonian formulation of N-bein gravity, though
both are symmetries of the original action. We have also shown that at the
Lagrangian level they do not form a group (\ref{dif10}). Indeed, it follows,
for example, from (\ref{eqn-diff}) that three invariances, rotation,
translation and diffeomorphism, are not independent. Only linearly independent
identities contribute to independent symmetries of an action \cite{Noether}.

In the Hamiltonian formulation of the first order EC gravity when using the
original variables, $e_{\tau\left(  \rho\right)  }$ and $\omega_{\nu\left(
\alpha\beta\right)  }$, diffeomorphism cannot be derived from the first class
constraints as a gauge symmetry (as we have shown in \cite{Report, 3D,
Darboux}). In addition, diffeomorphism has never been obtained for N-bein
gravity in the Hamiltonian formulation without some unjustified assumptions,
such as using a non-canonical change of variables, solving a first class
constraint, or fixing a \textquotedblleft gauge\textquotedblright%
\ \textit{before} the gauge symmetry was derived. In the Hamiltonian
formulation, both translational and diffeomorphism symmetries cannot
simultaneously be present as gauge symmetries and the only possibility of
reconciling these two symmetries as being gauge symmetries would be to find a
canonical transformation between the original variables, $e_{\gamma\left(
\rho\right)  }$ and $\omega_{\nu\left(  \alpha\beta\right)  }$, and some
others in terms of which we have diffeomorphism invariance.\footnote{Of
course, having solely spatial diffeomorphism for a covariant theory is not
acceptable. From the Hamiltonian point of view it guarantees that one has
performed a non-canonical change of variables \cite{Myths, affine-metric}.}
However, this is a faint possibility that such a transformation exists and it
is more likely that canonical transformations do not change a gauge symmetry
arising from the first class constraints \cite{FKK} (see also Section 5 of
\cite{affine-metric}).

What we can call a gauge symmetry\ is not to be determined by our choice and
all arguments based on \textquotedblleft convenience\textquotedblright\ or
\textquotedblleft custom\textquotedblright\ or referred to as being
\textquotedblleft physical\textquotedblright,\ \textquotedblleft
geometrical\textquotedblright\ are not acceptable. A \textit{mathematical}
criterion has to be developed to answer this question. It is dangerous and
purely speculative to draw any conclusion about a gauge symmetry without
justifying it with the well-defined mathematical procedure. In any case, the
gauge symmetries of a system (i.e. those invariances which follow from the
first class constraints present) must be found by following the well
prescribed procedure.

In the Lagrangian approach, there is nothing special in the question
\textquotedblleft diffeomorphism versus translational
invariance\textquotedblright, as many differential identities can be built
starting from basic identities and their combinations. We can construct even
different forms of translational invariance with the same, internal, index by
using, in addition to the naturally occurring $I^{\left(  \alpha\right)  }$
(\ref{eqn26}) (let us call this the \textquotedblleft zero
DI\textquotedblright, $I_{0}^{\left(  \alpha\right)  }$), other differential
identities; for example, from $I_{\mu}$ (\ref{eqn20})\textit{ }and $I^{\left(
\alpha\beta\right)  }$ (\ref{eqn14})\textit{ }we can construct $I_{1}^{\left(
\alpha\right)  }=e^{\mu\left(  \alpha\right)  }\omega_{\mu\left(  \gamma
\beta\right)  }I^{\left(  \gamma\beta\right)  }$, $I_{2}^{\left(
\alpha\right)  }=e^{\mu\left(  \alpha\right)  }I_{\mu}$, etc., and obtain from
them the corresponding transformations. Moreover, for DIs of the same
tensorial dimension we can consider linear combinations of them with real
coefficients $I^{\left(  \alpha\right)  }=%
{\displaystyle\sum\limits_{i=0}^{n}}
a_{i}I_{i}^{\left(  \alpha\right)  }$ which lead to any number of
translational symmetries. (Note that not all $I_{i}^{\left(  \alpha\right)  }$
are independent; for example, $I_{2}^{\left(  \alpha\right)  }=I_{0}^{\left(
\alpha\right)  }+I_{1}^{\left(  \alpha\right)  }$.) However, all these
symmetries cannot be generated simultaneously by the first class constraints
in the Hamiltonian formulation. Can all of them be gauge symmetries (but not
simultaneously)? Which of these \textquotedblleft
translations\textquotedblright\ form a group and exhibit group properties
with, for example, diffeomorphism and rotation? Translational invariance and
diffeomorphism together do not form a group (see (\ref{dif10})) but the
commutator of diffeomorphism and rotation does so (\ref{dif2}), and thus from
the Lagrangian point of view they form a group \cite{Noether}. This result
contradicts to the \textquotedblleft canonical\textquotedblright\ formulation
of the EC action where it is claimed that the PB among constraints responsible
for diffeomorphism and rotation are zero \cite{CNP, Henneaux, Maluf} which is
an additional indication that these results should be checked for correctness.

Finding translational invariance in N-bein gravity in any dimension is a very
simple task in the Lagrangian approach.\footnote{The exception is the three
dimensional case where totally antisymmetric $C^{\tau\left(  \rho\right)
\lambda\left(  \sigma\right)  \mu\left(  \alpha\right)  \nu\left(
\beta\right)  }$ in (\ref{eqn21}) (that has four indices internal or external)
is manifestly zero (see (\ref{eqn44}) or (\ref{eqn46})) and so it is necessary
in the Lagrangian approach to treat the three dimensional case separately in
order to find its translational invariance. This is also consistent with the
Hamiltonian formulation where in $D=3$ some peculiarities appear that are not
present in higher dimensions \cite{3D}. For peculiarities of sypersymmetric
extension in $D=3$ see \cite{Nicolai, WMN}.} The derivation of the
translational invariance of N-bein gravity in any dimension leads to a much
more general question: which symmetries of an action can be called
\textit{gauge} symmetries? The Hamiltonian approach allows us to address this
question and uniquely derive the gauge invariance for a particular formulation
of any theory. This is important, as in methods of quantization, such as in
path integral quantization using the Faddeev-Popov technique, it is necessary
to fix a gauge and one must know which symmetry (in the case of N-bein
gravity) has to be fixed (internal translation, and which one ($I_{0}^{\left(
\alpha\right)  },I_{1}^{\left(  \alpha\right)  },I_{2}^{\left(  \alpha\right)
},...$), or restrictions on possible coordinate transformations have to be
imposed). Some authors (see, e.g. \cite{DN}) have attempted to quantize
vierbein gravity using a gauge fixing which breaks rotational and
diffeomorphism invariances, without considering translational invariance.
However, any conclusion about the quantum behaviour of gravity should be made
only after we understand which gauge has to be fixed.\textit{ }

If the transformations (\ref{eqn16}), (\ref{eqn17}) and (\ref{eqn28}),
(\ref{eqn29}) are to be considered as the gauge transformations, then the
Hamiltonian formulation of the EC action has to have first class constraints
which lead to a gauge generator that generates these transformations without
any field dependent redefinition of gauge parameters.

It is possible that the Lagrangian approach can be used to single out which
symmetry is a gauge symmetry, that can be derived from the structure of first
class constraints of the Hamiltonian formulation of the same theory. This
conjecture is based on the following observation \cite{prep}: in field
theories (e.g. Maxwell, Yang-Mills, Einstein-Hilbert in its first or second
order formulations) the basic differential identities formed from the
derivatives of EDs lead to the invariance, exactly the same that follows from
the first class constraints in the Hamiltonian formulation of a theory. In the
EC action, the basic differential identities are (\ref{eqn14}) and
(\ref{eqn26}) and these lead to translation and rotation in the tangent space.
This is a pure Lagrangian result; we expect the same in the Hamiltonian
formulation. The equivalence of Hamiltonian and Lagrangian methods leads to
this conclusion and this equivalence also shows itself at some intermediate
steps; for example, the commutators of transformations following from basic
DIs and PBs among first class constraints responsible for such transformations
are in direct correspondence for known gauge theories \cite{prep}. For the EC
action the commutator of two rotations gives a rotation, (\ref{ct1}%
)/(\ref{ct3}), as when one computes the PB between two rotational constraints
in the Hamiltonian formulation \cite{3D, Report}; the commutator of rotation
and translation is proportional to translation only, (\ref{ct4})/(\ref{ct6}),
as we observe in the Hamiltonian formulation of the EC action \cite{3D,
Report}. The commutator of \ two translations is proportional to a translation
and a rotation; this must also be found for the PB of the corresponding
constraints in the Hamiltonian formulation. The complete Hamiltonian
formulation of the EC action is currently being investigated and the results
will be reported elsewhere. In addition, there is a correspondence that has to
be studied in which the commutators of transformations following from basic
DIs can provide a clue about the PB algebra of the first class constraints. If
in the commutator of two consecutive transformations the new parameter is
expressed through old parameters without involving fields (as in (\ref{ct2})
and (\ref{ct7})), then the PB between the corresponding constraints does not
have field dependent structure functions. If there are no derivatives of
parameters in the expression for the new parameter, then the corresponding PB
algebra is local (i.e. has no derivatives of delta functions). For example,
for the second order the Einstein-Hilbert action, the Lagrangian approach
leads to the known result (\ref{dif5}) (see \cite{Bergmann}), in which the new
parameter involves derivatives of old parameters and the corresponding PB
algebra of constraints is non-local in this case. These general properties of
the Lagrangian and Hamiltonian formulations of field theories \ should be
applicable to the EC action and this is exactly what we have already found by
considering the Lagrangian method in this paper and the preliminary results
following from the Hamiltonian formulation in \cite{3D, Report, Darboux}. This
provides an answer to Matschull's question posted for the EC theory:
\textquotedblleft what is a gauge symmetry and what is not\textquotedblright%
.\ The gauge, basic, invariances of the EC action which follow from its first
class constraints are translation and rotation in the tangent space, but not
diffeomorphism. We found that the differential identities also play important
role in analyzing consistency of coupling of gravity with other fields. The
results will be reported elsewhere.

\vspace{5mm}

\textbf{Acknowledgements}

\vspace{5mm}

We would like to thank A.M. Frolov, D.G.C. McKeon, and J. Nowak for helpful
discussion and reading the manuscript.

\appendix

\section{ $ABC$ properties}

Here we collect properties of the $ABC$ functions that were introduced in
considering the Hamiltonian formulation of N-bein gravity \cite{Report, 3D}.
They turn out to also be very useful in the Lagrangian approach.

These functions are generated by consecutive variation of the N-bein density%

\begin{equation}
\frac{\delta}{\delta e_{\nu\left(  \beta\right)  }}\left(  ee^{\mu\left(
\alpha\right)  }\right)  =e\left(  e^{\mu\left(  \alpha\right)  }e^{\nu\left(
\beta\right)  }-e^{\mu\left(  \beta\right)  }e^{\nu\left(  \alpha\right)
}\right)  =eA^{\mu\left(  \alpha\right)  \nu\left(  \beta\right)  },
\label{eqn40}%
\end{equation}

\begin{equation}
\frac{\delta}{\delta e_{\lambda\left(  \gamma\right)  }}\left(  eA^{\mu\left(
\alpha\right)  \nu\left(  \beta\right)  }\right)  =eB^{\lambda\left(
\gamma\right)  \mu\left(  \alpha\right)  \nu\left(  \beta\right)  },\text{ \ }
\label{eqn41}%
\end{equation}

\begin{equation}
\frac{\delta}{\delta e_{\tau\left(  \sigma\right)  }}\left(  eB^{\lambda
\left(  \gamma\right)  \mu\left(  \alpha\right)  \nu\left(  \beta\right)
}\right)  =eC^{\tau\left(  \sigma\right)  \lambda\left(  \gamma\right)
\mu\left(  \alpha\right)  \nu\left(  \beta\right)  },\quad... \label{eqn41a}%
\end{equation}

The first important property of these density functions is their total
antisymmetry: interchange of two indices of the same nature (internal or
external), e.g.%

\begin{equation}
A^{\nu\left(  \beta\right)  \mu\left(  \alpha\right)  }=-A^{\nu\left(
\alpha\right)  \mu\left(  \beta\right)  }=-A^{\mu\left(  \beta\right)
\nu\left(  \alpha\right)  } \label{eqn42}%
\end{equation}
with the same being valid for $B$, $C$, etc. In calculations presented here,
nothing is needed beyond $C$.

The second important property is their expansion using an external index%

\begin{equation}
B^{\tau\left(  \rho\right)  \mu\left(  \alpha\right)  \nu\left(  \beta\right)
}=e^{\tau\left(  \rho\right)  }A^{\mu\left(  \alpha\right)  \nu\left(
\beta\right)  }+e^{\tau\left(  \alpha\right)  }A^{\mu\left(  \beta\right)
\nu\left(  \rho\right)  }+e^{\tau\left(  \beta\right)  }A^{\mu\left(
\rho\right)  \nu\left(  \alpha\right)  }, \label{eqn43}%
\end{equation}

\begin{equation}
C^{\tau\left(  \rho\right)  \lambda\left(  \sigma\right)  \mu\left(
\alpha\right)  \nu\left(  \beta\right)  }=e^{\tau\left(  \rho\right)
}B^{\lambda\left(  \sigma\right)  \mu\left(  \alpha\right)  \nu\left(
\beta\right)  }-e^{\tau\left(  \sigma\right)  }B^{\lambda\left(
\alpha\right)  \mu\left(  \beta\right)  \nu\left(  \rho\right)  }%
+e^{\tau\left(  \alpha\right)  }B^{\lambda\left(  \beta\right)  \mu\left(
\rho\right)  \nu\left(  \sigma\right)  }-e^{\tau\left(  \beta\right)
}B^{\lambda\left(  \rho\right)  \mu\left(  \sigma\right)  \nu\left(
\alpha\right)  } \label{eqn44}%
\end{equation}
or an internal index%

\begin{equation}
B^{\tau\left(  \rho\right)  \mu\left(  \alpha\right)  \nu\left(  \beta\right)
}=e^{\tau\left(  \rho\right)  }A^{\mu\left(  \alpha\right)  \nu\left(
\beta\right)  }+e^{\mu\left(  \rho\right)  }A^{\nu\left(  \alpha\right)
\tau\left(  \beta\right)  }+e^{\nu\left(  \rho\right)  }A^{\tau\left(
\alpha\right)  \mu\left(  \beta\right)  }, \label{eqn45}%
\end{equation}

\begin{equation}
C^{\tau\left(  \rho\right)  \lambda\left(  \sigma\right)  \mu\left(
\alpha\right)  \nu\left(  \beta\right)  }=e^{\tau\left(  \rho\right)
}B^{\lambda\left(  \sigma\right)  \mu\left(  \alpha\right)  \nu\left(
\beta\right)  }-e^{\lambda\left(  \rho\right)  }B^{\mu\left(  \sigma\right)
\nu\left(  \alpha\right)  \tau\left(  \beta\right)  }+e^{\mu\left(
\rho\right)  }B^{\nu\left(  \sigma\right)  \tau\left(  \alpha\right)
\lambda\left(  \beta\right)  }-e^{\nu\left(  \rho\right)  }B^{\tau\left(
\sigma\right)  \lambda\left(  \alpha\right)  \mu\left(  \beta\right)  }.
\label{eqn46}%
\end{equation}

The third property involves their derivatives%

\begin{equation}
\left(  eA^{\nu\left(  \beta\right)  \mu\left(  \alpha\right)  }\right)
,_{\sigma}=\frac{\delta}{\delta e_{\lambda\left(  \gamma\right)  }}\left(
eA^{\nu\left(  \beta\right)  \mu\left(  \alpha\right)  }\right)
e_{\lambda\left(  \gamma\right)  ,\sigma}=eB^{\lambda\left(  \gamma\right)
\nu\left(  \beta\right)  \mu\left(  \alpha\right)  }e_{\lambda\left(
\gamma\right)  ,\sigma}\ , \label{eqn47}%
\end{equation}

\begin{equation}
\left(  eB^{\tau\left(  \rho\right)  \nu\left(  \beta\right)  \mu\left(
\alpha\right)  }\right)  ,_{\sigma}=\frac{\delta}{\delta e_{\lambda\left(
\gamma\right)  }}\left(  eB^{\tau\left(  \rho\right)  \nu\left(  \beta\right)
\mu\left(  \alpha\right)  }\right)  e_{\lambda\left(  \gamma\right)  ,\sigma
}=eC^{\tau\left(  \rho\right)  \lambda\left(  \gamma\right)  \nu\left(
\beta\right)  \mu\left(  \alpha\right)  }e_{\tau\left(  \rho\right)  ,\sigma
}\ . \label{eqn46a}%
\end{equation}

We also use the contraction of $B$ (\ref{eqn43}) with a covariant
$e_{\tau\left(  \lambda\right)  }:$%

\begin{equation}
e_{\tau\left(  \lambda\right)  }B^{\tau\left(  \rho\right)  \mu\left(
\alpha\right)  \nu\left(  \beta\right)  }=\delta_{\left(  \lambda\right)
}^{\left(  \rho\right)  }A^{\mu\left(  \alpha\right)  \nu\left(  \beta\right)
}+\delta_{\left(  \lambda\right)  }^{\left(  \alpha\right)  }A^{\mu\left(
\beta\right)  \nu\left(  \rho\right)  }+\delta_{\left(  \lambda\right)
}^{\left(  \beta\right)  }A^{\mu\left(  \rho\right)  \nu\left(  \alpha\right)
}. \label{eqn42a}%
\end{equation}

Upon using the antisymmetry of $B$ both in the external and internal indices
and the antisymmetry of $\omega$ in its internal indices leads to%

\begin{equation}
B^{\tau\left(  \rho\right)  \mu\left(  \alpha\right)  \nu\left(  \beta\right)
}\omega_{\mu\left(  \alpha\gamma\right)  }\omega_{\nu~~\sigma)}^{~(\gamma
}\omega_{\tau~~\beta)}^{~(\sigma}=0 \label{eqn50b}%
\end{equation}
and%

\begin{equation}
eB^{\tau\left(  \rho\right)  \mu\left(  \alpha\right)  \nu\left(
\beta\right)  }\omega_{\nu\left(  \alpha\beta\right)  ,\mu\tau}%
=0.\label{eqn50c}%
\end{equation}
\qquad\qquad\qquad\qquad\qquad A similar property for $C$ is%

\begin{equation}
C^{\lambda\left(  \gamma\right)  \nu\left(  \alpha\right)  \mu\left(
\rho\right)  \tau\left(  \sigma\right)  }e_{\lambda\left(  \gamma\right)
,\tau}e_{\nu\left(  \alpha\right)  ,\mu}=0. \label{eqn51}%
\end{equation}
The above properties considerably simplify calculations. The list of $ABC$
properties can be extended, but for our purpose the above relations are adequate.

\end{document}